\documentclass[aps,nofootinbib]{revtex4}
\usepackage{amsmath}
\usepackage{graphicx}
\usepackage{dcolumn}
\usepackage{bm}
\usepackage{amssymb}
\usepackage{latexsym}

\begin{document}
\begin{preprint}
 \qquad\qquad\qquad \qquad\qquad\qquad\qquad \qquad \qquad \qquad \qquad \qquad
\qquad \qquad \qquad \qquad \qquad \qquad \qquad \qquad \qquad
USTC-ICTS-07-03
\end{preprint}

\title{A String-Inspired Quintom Model Of Dark Energy}

\author{Yi-Fu Cai$^a$\footnote{caiyf@mail.ihep.ac.cn}, Mingzhe Li$^{b,c}$, Jian-Xin Lu$^d$, Yun-Song Piao$^e$, Taotao Qiu$^a$\footnote{qiutt@mail.ihep.ac.cn}, Xinmin Zhang$^a$}

\vspace{1.cm}
\address{$^a$Institute of High Energy Physics, Chinese Academy
of Sciences, P.O. Box 918-4, Beijing 100049, P. R. China
\vspace{.1in} \\
$^b$Institut f$\ddot{u}$r Theoretische Physik, Philosophenweg 16,
D-69120 Heidelberg, Germany
\vspace{.1in} \\
$^c$Fakult\"{a}t f\"{u}r Physik, Universit\"{a}t Bielefeld,
D-33615 Bielefeld, Germany
\vspace{.1in}\\
$^d$Interdisciplinary Center for Theoretical Study, University of
Science and Technology of China, Hefei, Anhui 230026, China
\vspace{.1in} \\
$^e$College of Physical Sciences, Graduate School of Chinese
Academy of Sciences, YuQuan Road 19A, Beijing 100049, China}

%%%%%%%%%%%%%%%%%%%%%%%%%%%%%%%%%%%%%%%%%%%%%%%%%%%%%%%%%%%%%%%%%%%%%%%%%%
\begin{abstract}

We propose in this paper a quintom model of dark energy with a
single scalar field $\phi$ given by the lagrangian ${\cal
L}=-V(\phi)\sqrt{1-\alpha^\prime\nabla_{\mu}\phi\nabla^{\mu}\phi
+\beta^\prime \phi\Box\phi}$. In the limit of
$\beta^\prime\rightarrow$0 our model reduces to the effective low
energy lagrangian of tachyon considered in the literature. We study
the cosmological evolution of this model, and show explicitly the
behaviors of the equation of state crossing the cosmological
constant boundary.
\end{abstract}

\maketitle

%%%%%%%%%%%%%%%%%%%%%%%%%%%%%%%%%%%%%%%%%%%%%%%%%%%%%%%%%%%%%%%%%%%%%%%%%%
\section{Introduction}

The current data from type Ia supernovae, cosmic microwave
background (CMB) radiation, and other cosmological
observations\cite{1998snia,Riess,Spergel,Seljak} have provided
strong evidences for a spatially flat and accelerated expanding
universe at the present time. Within the framework of the standard
cosmology, this acceleration can be understood by introducing a
mysterious component, dubbed dark energy (DE). The simplest
candidate for DE is a minor positive cosmological constant, but it
suffers from the fine-tuning and coincidence problems. As a possible
solution to these problems  various dynamical models of DE have been
proposed, such as quintessence. In the recent years with the
accumulated astronomical observational data it becomes possible to
probe the current and even early behavior of DE. Although the
current fits to the data in combination of the 3-year
WMAP\cite{3wmap}, the recently released 182 SNIa Gold
sample\cite{Riess06} and also other cosmological observational data
show the consistence of the cosmological constant, it is worth
noting that the dynamical DE models are not excluded and a class of
dynamical models with equation of state across $-1$, dubbed {\it
quintom}, is mildly favored (for recent references see e.g.
\cite{Lihong2006, Zhao2006}).

Theoretically it is a big challenge to the model building of the
quintom dark energy. The Ref.\cite{Quintom1} is the first paper
pointing out this challenge and showing explicitly the difficulty
of realizing $w$ crossing over $-1$ in the quintessence and
phantom like models.
%%%%%%%%%%%%%%%%%%%%%%%%%%%%%%%%%%%%%%%%%%%%%%%%%%%%%%%%%%%%%%%%%%%%%%%%%%%%%%%%%
%Although there is a special class of models to provide $w<-1$
%without quantum instability which named ghost condensation, this
%can not realize $w$ across $-1$ without instability because it
%requires infinitely long time for $w$ to become $-1$\cite{ghost3}.
%%%%%%%%%%%%%%%%%%%%%%%%%%%%%%%%%%%%%%%%%%%%%%%%%%%%%%%%%%%%%%%%%%%%%%%%%%%%%%%%%
In general with a single fluid or a single scalar field with a
lagrangian of form ${\cal L}={\cal
L}(\phi,\partial_\mu\phi\partial^\mu\phi)$ it has been proved
\cite{GBZhaoperturb} (see also \cite{Quintom1} and \cite{Caldwell2005})
that the dark
energy perturbation would be divergent as the equation of state
(EOS) $w$ approaches to $-1$. This ``no-go" theorem forbids the
dynamical models widely studied in the literature with a single
scalar field such as quintessence, phantom and k-essence to make
the EOS cross over the cosmological constant boundary. The quintom
scenario of dark energy is designed to understand the nature of
dark energy with $w$ across $-1$. The quintom models of dark
energy differ from the quintessence, phantom and k-essence and so
on in the determination of the cosmological evolution and the fate
of the universe.

To realize a viable quintom scenario of dark energy it needs to
introduce extra degree of freedom to the conventional theory with a
single fluid or a single scalar field. The first model of quintom
scenario of dark energy is given in Ref.\cite{Quintom1} with two
scalar fields. This model has been studied in detail later on. In
the recent years there have been active studies on models of quintom
like dark energy such as models with high derivative term\cite{lfz},
vector field\cite{Wei}, or even extended theory of
gravity\cite{extendgravity} and so on, see e.g. \cite{Quintomsum}.

In this paper, we propose a new type of quintom model inspired by
the string theory. We will demonstrate in this paper our model can
realize the equation of state crossing $-1$ naturally. This paper
is organized as follows. In section 2 we present our model and
study its properties especially on the conditions required for the
model parameters when $w$ crosses over $-1$. By solving
numerically the model we will study the evolution of the equation
of state. Moreover, we study the stability property of this model by
investigating the quadratic perturbations. The section 3 is the summary
of our paper.

\section{Our Model}

Our model is given by the following action\footnote{We have adopted
signature $(+, -, -, -)$ in this paper.}
\begin{eqnarray}\label{actionorigin}
S=\int d^4x\sqrt{-g}\left[-V(\phi)\sqrt{1-{\alpha}^\prime
\nabla_{\mu}\phi\nabla^{\mu}\phi+{\beta}^\prime\phi\Box\phi}\right].
\end{eqnarray}
This model generalizes the usual``Born-Infeld-type" action for the
effective description of tachyon dynamics by adding a term $\phi
\Box \phi$ to the usual $\nabla_\mu \phi\nabla^\mu \phi $ in the
square root. It is known that the 4D effective action of tachyon
dynamics to the lowest order in $ \nabla_\mu\phi\nabla^\mu \phi$
around the top of the tachyon potential can be obtained by the
stringy computations for either a D3 brane in bosonic theory
\cite{Gerasimov:2000zp, Kutasov:2000qp} or a non-BPS D3 brane in
supersymmetric theory \cite{Kutasov:2000aq}. To this order, we
have no need to include the operator $\phi \Box \phi$ in the
action since it is equivalent to the usual $\nabla_\mu
\phi\nabla^\mu \phi $ term. However, we cannot exclude its
existence in an action such as the ``Born-Infeld-type" one when
incorporating an infinite number of higher derivative terms since
now the two terms are in general different dynamically. For
example, we cannot simply replace the $\phi \Box \phi$ term in the
action (\ref{actionorigin}) by the $\nabla_\mu \phi\nabla^\mu \phi
$. Further this term in the above generalized action has new
cosmological consequence as will be shown in this paper. Including
an infinite number of higher derivative terms that has a
significant cosmological consequence has also been discussed in
the context of p-adic string recently in \cite{Barnaby:2006hi}.
The two parameters $\alpha'$ and $\beta'$ in (\ref{actionorigin})
can also be made arbitrary when the background flux is turned on
\cite{Mukhopadhyay:2002en}. Without further reasoning, we will
take the action (\ref{actionorigin}) as our starting point of
phenomenological study.

As will be demonstrated, the $\beta'$ term in (\ref{actionorigin})
is crucial to realize the $w$ across $-1$. There we have defined
$\alpha^\prime = \alpha/ M^4$ and $\beta^\prime = \beta/ M^4$ with
$\alpha$ and $\beta$ being the dimensionless parameters
respectively, and $M$ an energy scale used to make the ``kinetic
energy terms" dimensionless. $V(\phi)$ is the potential of scalar
field $\phi$ (e.g., a tachyon) with dimension of ${\rm [mass]}^4$
with an expected tachyon potential behavior in general, i.e.,
bounded and reaching its minimum asymptotically, unless
specifically stated. Note that,
$\Box=\frac{1}{\sqrt{-g}}\partial_{\mu}\sqrt{-g}g^{\mu\nu}\partial_{\nu}$,
therefore, in (\ref{actionorigin}) the terms
$\nabla_\mu\phi\nabla^\mu\phi$ and $\phi\Box\phi$ both involve two
fields and two derivatives.

The model with operator $\phi\Box\phi$ for the realization of $w$
crossing $-1$ has been proposed in \cite{lfz}. However in general
for a model with lagrangian as a sum of operators with a
polynomial function of the scalar field $\phi$ and its
derivatives, the operator $\phi\Box\phi$ can be rewritten as a
total derivative term which makes no contribution after
integration and a term which renormalizes the canonical kinetic
term as discussed above. So if one considers a renormalizable
lagrangian, the operator $\phi\Box\phi$ will not be included.
Ref.\cite{lfz} considered a dimension-6 operator as $(\Box
\phi)^2$. However in the present model, the operator
$\phi\Box\phi$ appears at the same order as the operator
$\nabla_\mu\phi\nabla^\mu\phi$ does in the ``Born-Infeld-type"
action. As discussed above, this model appears more natural than
the one used in \cite{lfz}.

With (\ref{actionorigin}) we obtain the equation of motion of the
scalar field $\phi$,
\begin{eqnarray}\label{Eqofmbasic}
\frac{\beta}{2}\Box(\frac{V\phi}{f})+
\alpha\nabla_{\mu}(\frac{V\nabla^{\mu}\phi}{f})+M^4V_{\phi}f+\frac{\beta
V}{2f}\Box\phi=0~,
\end{eqnarray}
where $f=\sqrt{1-{\alpha}^\prime
\nabla_{\mu}\phi\nabla^{\mu}\phi+{\beta}^\prime\phi\Box\phi}$ and
$V_\phi=dV /d\phi$. Following the convention in Ref.\cite{lfz}, the
energy-momentum tensor $T^{\mu\nu}$ is given by the standard
definition: $\delta_{g_{\mu\nu}}S\equiv-\int d^4 x
\frac{\sqrt{-g}}{2}T^{\mu\nu}\delta g_{\mu\nu}$,
\begin{eqnarray}\label{emtensor}
T^{\mu\nu}&=&g^{\mu\nu}[V(\phi)f+\nabla_{\rho}(\psi\nabla^{\rho}\phi)]+
\frac{\alpha}{M^4}\frac{V(\phi)}{f}\nabla^\mu\phi\nabla^\nu\phi-\nabla^\mu\psi\nabla^\nu\phi-\nabla^\nu\psi\nabla^\mu\phi~,
\end{eqnarray}
where $\psi\equiv\frac{\partial{\cal
L}}{\partial\Box\phi}=-\frac{V\beta\phi}{2M^4f}$.

For a flat Friedmann-Robertson-Walker (FRW) universe and a
homogenous scalar field $\phi$, the equation of motion in
(\ref{Eqofmbasic}) can be solved equivalently by the following two
equations
\begin{eqnarray}\label{EqofmTT}
\ddot\phi+3H\dot\phi&=&\frac{\beta
V^2\phi}{4M^4\psi^2}-\frac{M^4}{\beta\phi}+\frac{\alpha}{\beta\phi}\dot\phi^2~,\\
\ddot\psi+3H\dot\psi&=&(2\alpha+\beta)(\frac{M^4\psi}{\beta^2\phi^2}-\frac{V^2}{4M^4\psi})-(2\alpha-\beta)\frac{\alpha\psi}{\beta^2\phi^2}\dot\phi^2-\frac{2\alpha}{\beta\phi}\dot\psi\dot\phi-\frac{\beta
V\phi}{2M^4\psi}V_{\phi}~,
\end{eqnarray}
where we have made use of the $\psi$ as defined before and the first
equation above is just the defining equation for $\psi$ in terms of
$\phi$ and its derivatives. $H = \dot a /a$ is the Hubble parameter.

 One can see from equations above the
$\beta$ term plays a role in determining the evolution of the scalar
field $\phi$. We can read the energy density from (\ref{emtensor})
as
\begin{eqnarray}\label{Energydensity}
\rho=Vf+\frac{d}{a^3dt}(a^3\psi\dot\phi)+\frac{\alpha}{M^4}\frac{V(\phi)}{f}\dot\phi^2-2\dot\psi\dot\phi~,
\end{eqnarray}
and similarly the pressure
\begin{eqnarray}\label{Pressure}
p=-Vf-\frac{d}{a^3dt}(a^3\psi\dot\phi)~.
\end{eqnarray}

With the Friedman equations $H^2=\frac{8\pi G}{3}\rho$ and $\dot
H=-4\pi G(\rho+p)$, we now study the cosmological evolution of
equation of state for the present model. Given $w=p/\rho$, we have
$\rho+p=(1+w)\rho$. To explore the possibility of the $w$ across
$-1$, we need to check if $\frac{d}{dt}(\rho+p)\neq0$ can be held
when $w \rightarrow -1$.

Using (\ref{Energydensity}) and (\ref{Pressure}) as well as  making
use of the defining equation for $\psi$, we have
\begin{eqnarray}\label{crosscondition1}
\rho+p=\frac{\alpha
V(\phi)\dot\phi^2}{M^4f}+\beta\dot\phi\frac{d}{dt}[\frac{\phi
V(\phi)}{M^4f}]~,
\end{eqnarray}
and from which,
\begin{eqnarray}\label{x}
\frac{d}{dt}(\rho+p)&=&\frac{\alpha}{M^4f^2}\{f\frac{d}{dt}
[V(\phi)\dot\phi^2]-\frac{df}{dt}
[\dot\phi^2{V}(\phi)]\}+\beta\ddot\phi\frac{d}{dt}[\frac{\phi
V(\phi)}{M^4f}]+\beta\dot\phi\frac{d^2}{dt^2}[\frac{\phi
V(\phi)}{M^4f}]~.
\end{eqnarray}
Eq. (\ref{crosscondition1}) implies that we have either
(i)$\dot\phi=0$ or (ii)$-\frac{\alpha
V(\phi)}{f}\dot\phi=\beta\frac{d}{dt}[\frac{\phi V(\phi)}{f}]$ when
$w \rightarrow -1$.

Let us assume $\dot\phi=0$ first when $w \to -1$. With this and Eq.
(\ref{x}), we have
$\frac{d}{dt}(\rho+p)=\beta\ddot\phi\frac{d}{dt}[\frac{\phi
V(\phi)}{M^4f}]=-\frac{V(\phi)}{2M^8f^3}\beta^2\phi^2\ddot\phi\frac{d}{dt}\Box\phi$.
Therefore the conditions for having the $w$ across over $-1$ are
(i-1)$\phi\neq0$; (i-2)$\ddot\phi\neq0$;
(i-3)$\frac{d}{dt}\Box\phi\neq0$ in addition to the $\dot \phi = 0$.
Since the characteristic behavior of $V(\phi)$ is to have a finite
maximum value at a finite $\phi$ and to reach its (exponentially)
vanishing minimum asymptotically\footnote{This is the behavior of a
tachyon potential.} where one expects $\ddot \phi = 0$ and
$\frac{d}{dt}\Box\phi = 0$ but a non-zero $\dot \phi$, so realizing
the above crossing over conditions must happen before reaching the
potential minimum asymptotically. This implies once the crossing
over conditions are met, the field $\phi$ must continue to run away
as it should be since we have $\ddot \phi \neq 0$.

Let us turn to the second case (ii) when $w \to -1$. We now have
$-\frac{\alpha V\dot\phi}{f}=\beta\frac{d}{dt}(\frac{\phi
V}{f})=\beta\dot\phi(\frac{V}{f})+\beta\phi\frac{d}{dt}(\frac{V}{f})$
which is equivalent to saying $Y \equiv (\alpha + \beta) \frac{V}{f}
\dot\phi + \beta\phi\frac{d}{dt}(\frac{V}{f}) = 0$. For now, we have
$\dot\phi \neq 0$. Then the above can be further expressed as
$\frac{d}{dt}[(\alpha+\beta)\ln\phi+\beta \ln(\frac{V}{f})]=0$ if
$\phi \neq 0$.\footnote{We can always choose $\phi \neq 0$ and the
$\phi = 0$ will give a degenerate case which will not be considered
here.} Then from (\ref{x}), the further condition for crossing over
is $\frac{d}{dt}(\rho+p)\neq0$ which implies $\dot Y \neq 0$ or
$\frac{d^2}{dt^2}[(\alpha+\beta)\ln\phi+\beta
\ln(\frac{V}{f})]\neq0$. In summary,  the following conditions are
required for the crossing over: ii-1)$Y = 0$ and ii-2) $\dot Y \neq
0$ in addition to the $\dot\phi \neq 0$ and $\phi \neq 0$.
%$\footnote{We use the $Y$ defined in the context not a simpler one
%such as $Y = \phi^{\alpha + \beta} (\frac{V}{f})^\beta$ because the
%$Y$ is well defined asymptotically and is useful for arguing the
%crossing over.}.
 Given the definition of the above $Y$ and the
characteristic behavior of $V(\phi)$ discussed previously, one
expects that both $Y$ and $\dot Y$ vanish asymptotically while $\dot
\phi$ can reach a non-vanishing value. This implies that the
crossing over if occurring at all must occur before the $V(\phi)$
reaches its minimum asymptotically as anticipated.
% In general, one
%expects that realizing the present case with $\dot\phi \neq 0$ is
%easier than the previous one with $\dot\phi = 0$ at the crossing
%over.

Before we demonstrate numerically that the crossing over of the
present model can indeed be realized using specific examples, let us
remark one salient feature of the present model that the $\beta$
term in the present model (\ref{actionorigin}) is the key for
realizing the crossing over. One can check from
(\ref{crosscondition1}) that if $\beta = 0$ the $\omega \to - 1$ is
possible only for $\dot\phi = 0$. Then from (\ref{x}), we will have
$\frac{d}{dt}(\rho+p) = 0$, the impossibility of crossing over.

\begin{figure}[htbp]
\includegraphics[scale=0.8]{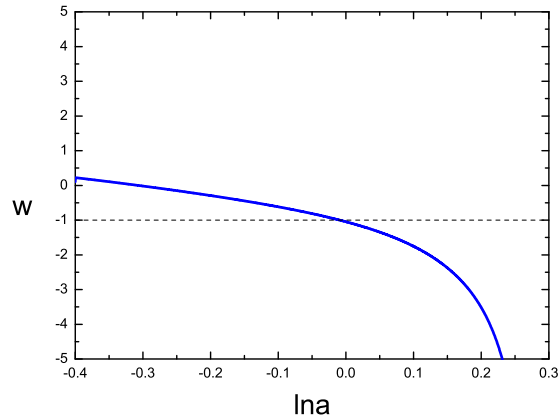}
\caption{Plot of the evolution of the EOS as a function of $\ln
a$. In the numerical calculation, we have taken $V(\phi)=V_0
e^{-\lambda\phi^2}$ with $V_0=0.8$, $\lambda=1$, $\alpha=1$, and
$\beta=-0.8$. For the initial conditions we choose $\phi_i=0.9$,
$\dot\phi_i=0.6$, $(\Box\phi)_i=\frac{d}{dt}(\Box\phi)_i=0$.}
\label{fig:1bigrip}
\end{figure}

The analysis above shows various possibilities of our model in
realizing the EOS $w$ crossing $-1$. Now we consider some specific
examples for numerical calculations of the evolution of the EOS. In
Figure \ref{fig:1bigrip} we take $V(\phi)=V_0 e^{-\lambda\phi^2}$
motivated by string theory and plot the behavior of EOS. In the
numerical calculations we have normalized the values of the scalar
field and $V_0$, respectively, by the energy scale $M$. In Figure
\ref{fig:1bigrip} our model predicts the EOS crossing $-1$ during
the evolution and a big-rip singularity for the fate of the
universe. Numerically we have checked that $\dot \phi = 0$ when $w$
crosses over the cosmological constant boundary.

\begin{figure}[htbp]
\includegraphics[scale=0.8]{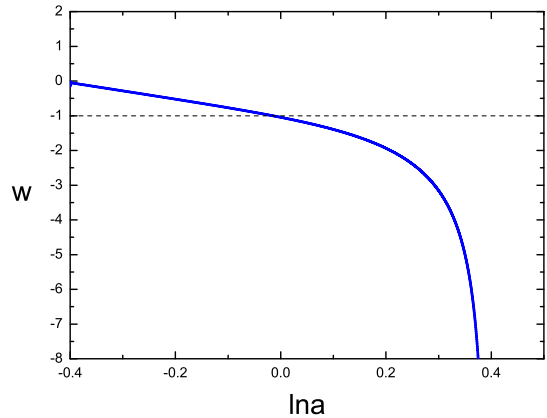}
\caption{Plot of the evolution of the EOS as a function of $\ln a$.
In the numerical calculation we take
$V(\phi)=\frac{V_0}{e^{\lambda\phi}+e^{-\lambda\phi}}$, and
$V_0=0.5$. For the model parameters we choose $\lambda=1$,
$\alpha=1$, and $\beta=-0.8$. For the initial conditions we take
$\phi_i=0.9$, $\dot\phi_i=0.6$,
$(\Box\phi)_i=\frac{d}{dt}(\Box\phi)_i=0$.}\label{fig:2}
\end{figure}

In Figure \ref{fig:2} we take a different potential for numerical
calculations. One can see that the EOS crosses over $-1$ during the
evolution. When we take $\beta$ positive Figure 3 shows the EOS
starts with $w<-1$, crosses over $-1$ into the region of $w>-1$,
then transits again to $w<-1$.

\begin{figure}[htbp]
\includegraphics[scale=0.8]{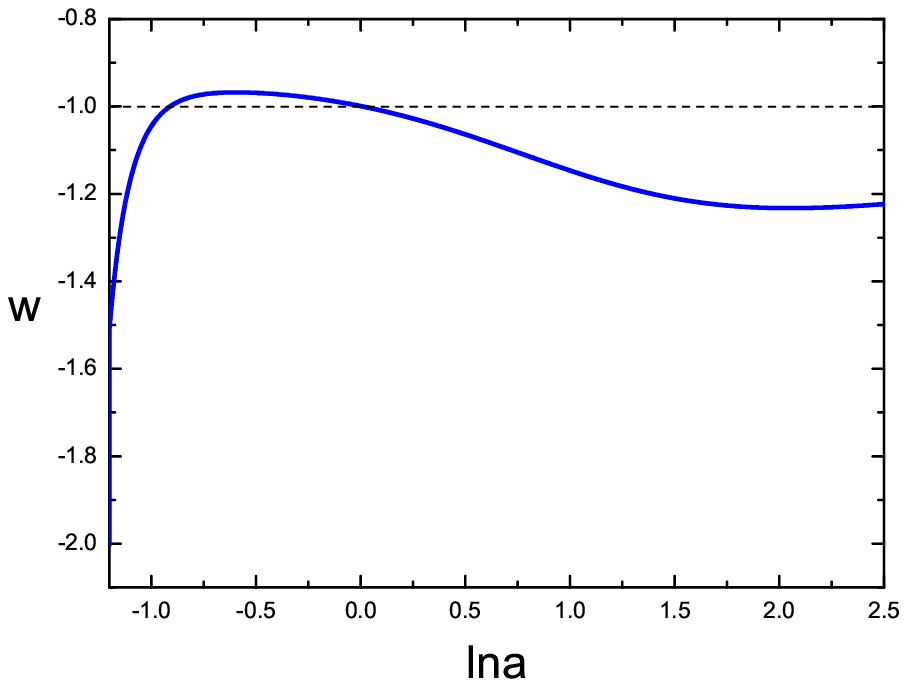}
\caption{Plot of the evolution of the EOS as a function of $\ln a$.
In the numerical calculation we take
$V(\phi)=\frac{V_0}{e^{\lambda\phi}+e^{-\lambda\phi}}$, and
$V_0=0.5$. For the model parameters we choose $\lambda=1$,
$\alpha=1$, and $\beta=0.8$. For the initial conditions we take
$\phi_i=0.9$, $\dot\phi_i=0.6$,
$(\Box\phi)_i=\frac{d}{dt}(\Box\phi)_i=0$.}\label{fig:3}
\end{figure}

%The potentials used for the numerical calculations in Figures 1-3
%are well motivated by the string theory. However as a
%phenomenological study of our model as a quintom dark energy we in
%Figure 4 plot the evolution of the EOS $w$ with a potential which is
%a sum of $e^{\lambda \phi}$ and $e^{- \lambda \phi}$. This type of
%potential does not have the general behavior of the tachyon
%potential, however has been used for the study of phenomenological
%models of dark energy. One can see from this Figure that the EOS
%evolves from the region where $w> -1$ to $w<-1$, and stays there
%for a period of time then comes back to $w>-1$. At late time our
%model gives rise to the de-sitter phase.

%\begin{figure}[htbp]
%\includegraphics[scale=0.8]{fig4.eps}
%\caption{Plot of the evolution of the EOS as a function of $\ln a$.
%In the numerical calculation we take
%$V(\phi)=V_0(e^{\lambda\phi}+e^{-\lambda\phi})$, and $V_0=0.5$. For
%the model parameters we choose $\lambda=1$, $\alpha=1$, and
%$\beta=-1.2$. For the initial conditions we take $\phi_i=0.9$,
%$\dot\phi_i=0.6$, $(\Box\phi)_i=\frac{d}{dt}(\Box\phi)_i=0$.}
%\label{fig:3}
%\end{figure}
%############################################################

Now we study the stability property of this model by considering the
quadratic perturbations.
Consider a small perturbation $\pi$
around the background $\phi$,
\begin{eqnarray}\label{phidecompose}
\phi\rightarrow\phi(t)+\pi(t,{\bf x})~,
\end{eqnarray}
where the background field $\phi$ in the FRW cosmology is
spatially homogenous. Working with the action in
(\ref{actionorigin}) after a shift of the field in
(\ref{phidecompose}) and a tedious calculation we obtain the terms
for the quadratic perturbations
\begin{eqnarray}
^{(2)}S &\sim&\int
d^4x\sqrt{-g}~[(\frac{\alpha'}{2}Vf^{-1}+\frac{\alpha'^2}{2}Vf^{-3}\dot\phi^2+\frac{\beta'}{2}Vf^{-1}+\frac{\beta'}{2}\phi
V_\phi f^{-1}-\frac{\beta'^2}{4}\phi\Box\phi
Vf^{-3})\dot\pi^2\nonumber\\
&~&-(\frac{\alpha'}{2}Vf^{-1}+\frac{\beta'}{2}Vf^{-1}+\frac{\beta'}{2}\phi
V_\phi f^{-1}-\frac{\beta'^2}{4}\phi\Box\phi
Vf^{-3})(\nabla\pi)^2\nonumber\\
&~&+...~].
\end{eqnarray}

\begin{figure}[htbp]
\includegraphics[scale=0.7]{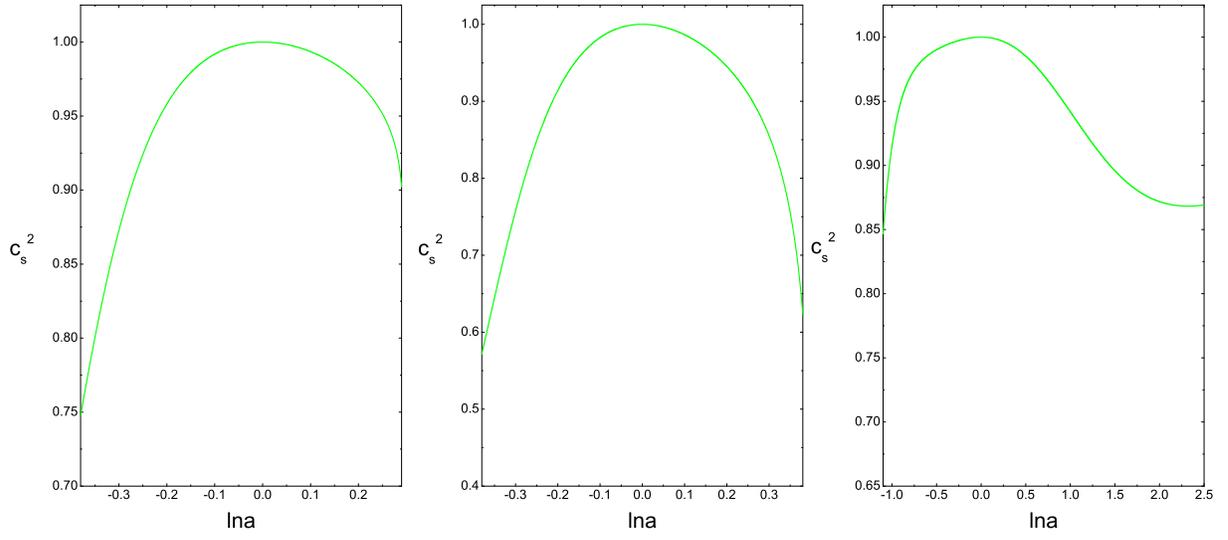}
\caption{Plots of the sound speeds (from left to right) for the three
models considered in this
paper for the numerical calculations shown in Figure 1-3.} \label{fig:cs}
\end{figure}

\begin{figure}[htbp]
\includegraphics[scale=0.7]{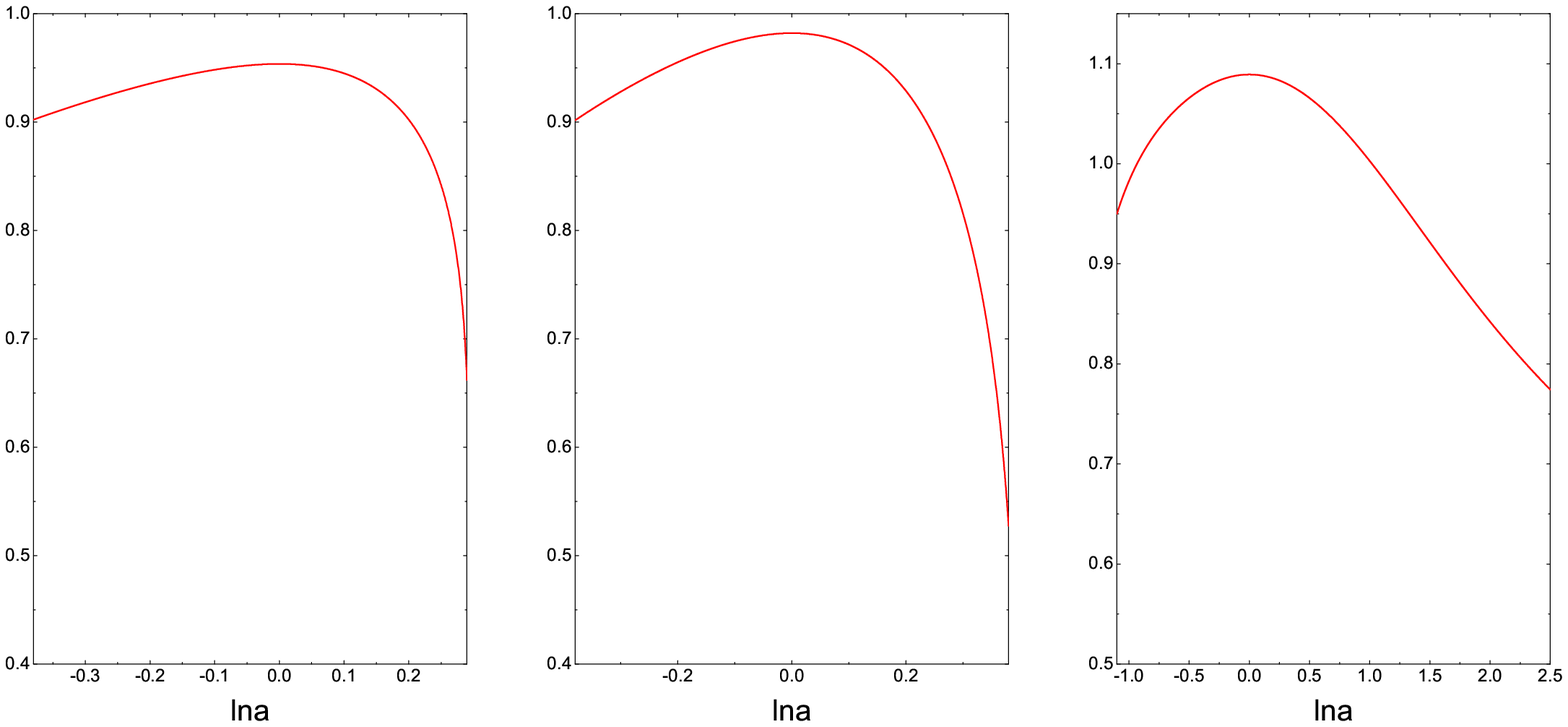}
\caption{Plots of the coefficient of $\dot\pi^2$ (from left to
right) for the three models considered in this paper for the
numerical calculations shown in Figure 1-3.} \label{fig:kin}
\end{figure}

Interestingly we notice that due to the positivity of the term
$\frac{\alpha'^2}{2}Vf^{-3}\dot\phi^2$ in this model if the
coefficient of $(\nabla\pi)^2$ is positive, the term in front of
$\dot\pi^2$ guarantee to be positively valued. The sound speed
characterizing the stability property of the perturbations is
given by
\begin{eqnarray}
c_s^2=\frac{\frac{\alpha'}{2}Vf^{-1}+\frac{\beta'}{2}Vf^{-1}+\frac{\beta'}{2}\phi
V_\phi f^{-1}-\frac{\beta'^2}{4}\phi\Box\phi
Vf^{-3}}{\frac{\alpha'}{2}Vf^{-1}+\frac{\alpha'^2}{2}Vf^{-3}\dot\phi^2+\frac{\beta'}{2}Vf^{-1}+\frac{\beta'}{2}\phi
V_\phi f^{-1}-\frac{\beta'^2}{4}\phi\Box\phi Vf^{-3}}~.
\end{eqnarray}
If the $c_s^2$ lies in the range between $0$ and $1$
%########
and the coefficient of $\dot\pi^2$ keeps positive,
%########
our model will be
stable. In Figure \ref{fig:cs}
%########
and Figure \ref{fig:kin}
%########
we plot the $c_s^2$ and the coefficients of $\dot\pi^2$
respectively for the models above in the figures 1-3. From these
figures one can see for these models $c_s^2$ are in the range of
0-1 and the coefficients of $\dot\pi^2$ are positive, so our
models are stable.

%############################################################

\section{Conclusion and Discussion}

The current cosmological observations indicate the possibility
that the acceleration of the universe is driven by dark energy
with EOS across $-1$, which if confirmed further in the future
will challenge the theoretical model building of the dark energy.
In this paper we have proposed a string-inspired model of dark
energy through modifying the usual effective ``Born-Infeld-type"
description of tachyon dynamics. As shown in the present work,
this modification by including a $\beta$ term in the action
(\ref{actionorigin}) is the key for the EOS crossing $-1$ during
the evolution.\footnote{The $\beta\rightarrow $0 limit reduces the
present model to the effective low energy Lagrangian of
tachyon\cite{Sentachyon} which has been considered to be a
candidate for dark energy in the literature\cite{TachyonDE}. This
type of models for dark energy in the absence of the $\beta$ term,
however will not be possible to realize the quintom scenario as
shown in the present work.} Compared to other models with $w$
across $-1$ in the literature so far the present one is also
economical in the sense that it involves a single scalar field
such as a tachyon and has a motivation inspired from string theory
consideration since the new features of this model for the dark
energy could also be present in the realistic string theory if the
whole tower of the higher derivative terms is fully included in
the effective action.

\acknowledgments

We thank Bo Feng, Gongbo Zhao, Hong Li and Junqing Xia for useful
discussions. This work is supported in part by National Natural
Science Foundation of China under Grant Nos. 90303004, 10533010 and
19925523. The author M.L. would like to acknowledge the support by
Alexander von Humboldt Foundation. JXL acknowledges support by
grants from the Chinese Academy of Sciences and grants from the NSF
of China with Grant Nos: 10588503 and 10535060.

\end{document}